\documentclass[%
reprint,
superscriptaddress,
 amsmath,amssymb,
 aps,
pra,
]{revtex4-2}

\usepackage{color}
\usepackage{pst-plot}
\usepackage{pst-optexp}
\usepackage{graphicx}
\usepackage{dcolumn}
\usepackage{bm}
\usepackage{hyperref}
\usepackage{ulem}

\newcommand{\bra}[1]{\langle #1|}
\newcommand{\ket}[1]{|#1\rangle}

\begin{document}

\preprint{APS/123-QED}

\title{Enhanced nonlinear interferometry via seeding}



\author{J.~Fl\'orez}
\email{j.florez-gutierrez@imperial.ac.uk}
\affiliation{Department of Physics, Blackett Laboratory, Imperial College London, South Kensington Campus, London SW7 2AZ, United Kingdom}
\author{E.~Pearce}
\affiliation{Department of Physics, Blackett Laboratory, Imperial College London, South Kensington Campus, London SW7 2AZ, United Kingdom}
\author{N.~R.~Gemmell}
\affiliation{Department of Physics, Blackett Laboratory, Imperial College London, South Kensington Campus, London SW7 2AZ, United Kingdom}
\author{Y.~Ma}
\affiliation{Department of Physics, Blackett Laboratory, Imperial College London, South Kensington Campus, London SW7 2AZ, United Kingdom}
\author{G. Bressanini}
\affiliation{Department of Physics, Blackett Laboratory, Imperial College London, South Kensington Campus, London SW7 2AZ, United Kingdom}
\author{C.~C.~Phillips}
\affiliation{Department of Physics, Blackett Laboratory, Imperial College London, South Kensington Campus, London SW7 2AZ, United Kingdom}
\author{R.~F.~Oulton}
\affiliation{Department of Physics, Blackett Laboratory, Imperial College London, South Kensington Campus, London SW7 2AZ, United Kingdom}
\author{A.~S.~Clark}
\affiliation{Department of Physics, Blackett Laboratory, Imperial College London, South Kensington Campus, London SW7 2AZ, United Kingdom}
\affiliation{Quantum Engineering Technology Labs, H. H. Wills Physics Laboratory and Department of Electrical and Electronic Engineering, University of Bristol, BS8 1FD, United Kingdom}


\date{\today}

\begin{abstract}
We analyse a nonlinear interferometer, also known as an SU(1,1) interferometer, in the presence of internal losses and inefficient detectors. To overcome these limitations, we consider the effect of seeding one of the interferometer input modes with either a number state or a coherent state. We derive analytical expressions for the interference visibility, contrast, phase sensitivity, and signal-to-noise ratio, and show a significant enhancement in all these quantities as a function of the seeding photon number. For example, we predict that, even in the presence of substantial losses and highly inefficient detectors, we can achieve the same quantum-limited phase sensitivity of an unseeded nonlinear interferometer by seeding with a few tens of photons. Furthermore, we observe no difference between a number or a coherent seeding state when the interferometer operates in the low-gain regime, which enables seeding with an attenuated laser. Our results expand the nonlinear interferometry capabilities in the field of quantum imaging, metrology, and spectroscopy under realistic experimental conditions.
\end{abstract}

\maketitle
\section{Introduction}
Nonlinear interferometry has become an active research field in recent years thanks to its demonstrated and potential applications, which include imaging~\cite{Lemos14,Kviatkovsky20,Gilaberte21}, spectroscopy~\cite{Kalashnikov16,Paterova20,Kaufmann22}, optical coherence tomography~\cite{Paterova18,Vanselow20,Machado20,Rojas21}, holography~\cite{Topfer22}, multiphoton absorption measurements~\cite{Panahiyan22}, and sub-shot-noise phase sensitivity~\cite{Linnemann16,Manceau17}. One of the reasons for this growing interest in nonlinear interferometers, also known as SU(1,1) interferometers, is that the wavelengths involved in the interference process may belong to different spectral regions. For example, one wavelength can be in the visible or near-infrared (near-IR) region, where detection technology is well developed, while the correlated wavelength is in the mid-infrared (mid-IR), where detectors are noisy and less readily available. This wavelength versatility allows one to probe a sample with mid-IR light while recording the interference pattern in the visible region using off-the-shelf detectors \cite{Lemos14,Kviatkovsky20,Gilaberte21,Kalashnikov16,Paterova20,Kaufmann22,Paterova18,Vanselow20,Machado20,Rojas21,Topfer22}. Another interesting feature of nonlinear interferometers is their intrinsic quantum nature that allows phase measurements with sensitivities beyond the classical limit~\cite{Yurke86,Linnemann16,Manceau17}. This sub-shot-noise phase sensitivity plays an important role in the field of quantum metrology, with applications in the detection of gravitational waves and the dark matter axion field~\cite{Caves20}.

In practice, nonlinear interferometers are drastically affected by internal losses and detector imperfections. For instance, visibilities as low as $\sim$40\% are typical in this kind of interferometer~\cite{Paterova20,Kaufmann22,Vanselow20}, although the latter can be improved up to $\sim$90\% with either enhanced photon detectors and cameras~\cite{Paterova18,Gilaberte21}, or bright correlated-light sources~\cite{Manceau17,Frascella19,Machado20}. Other strategies to overcome the limitations of nonlinear interferometers include pumping the nonlinear crystals in the interferometer with unbalanced powers~\cite{Manceau17,Giese17,Gemmell22}, and seeding the interferometer input modes~\cite{Plick10,Marino12,Anderson17,Cardoso18,Michael21}. This last strategy may be useful in scenarios where unbalancing the pump powers is impractical, or when enhanced detectors or bright correlated-light sources are unavailable.

In this paper, we present a comprehensive study on the effect of seeding a non-degenerate nonlinear interferometer. We include internal losses and inefficient detectors in our theoretical calculations, as well as unbalanced parametric gains for completeness. We consider the cases of number and coherent state seeding of one of the interferometer input modes and provide analytical expressions for visibility, contrast, phase sensitivity, and signal-to-noise ratio (SNR). In extension to previous works where seeding was investigated~\cite{Plick10,Cardoso18,Michael21}, we describe how the interferometer properties are enhanced as a function of the seeding photon number. We focus on realistic experimental conditions, including substantial internal losses and highly inefficient detectors for a nonlinear interferometer working in the spontaneous or low-gain regime. Our results are especially relevant in the field of quantum metrology, where we predict an enhanced phase sensitivity with a few tens of seeding photons compared to an unseeded interferometer. Seeding may also play a relevant role in the field of spectroscopy, where the seeding photons can stimulate a particular spatial-spectral mode in the nonlinear interferometer to get enhanced spectral information about a low-concentration sample.

This paper is organised as follows. In Sec.~\ref{s:NLImodel} we provide a general description of a lossy and detection-inefficient nonlinear interferometer in a compact matrix form. We then describe the interference visibility and contrast in Sec.~\ref{s:VisibilityContrast}, and the phase sensitivity and SNR in Sec.~\ref{s:phasesensitivity}, as a function of the seeding photon number. Finally, in Sec.~\ref{s:conclusions}, we summarise our findings and discuss the similarities of seeding with a number versus a coherent state.

\section{Nonlinear interferometer model}\label{s:NLImodel}
\begin{figure}[b!]
	\centering
	\includegraphics[width=0.45\textwidth]{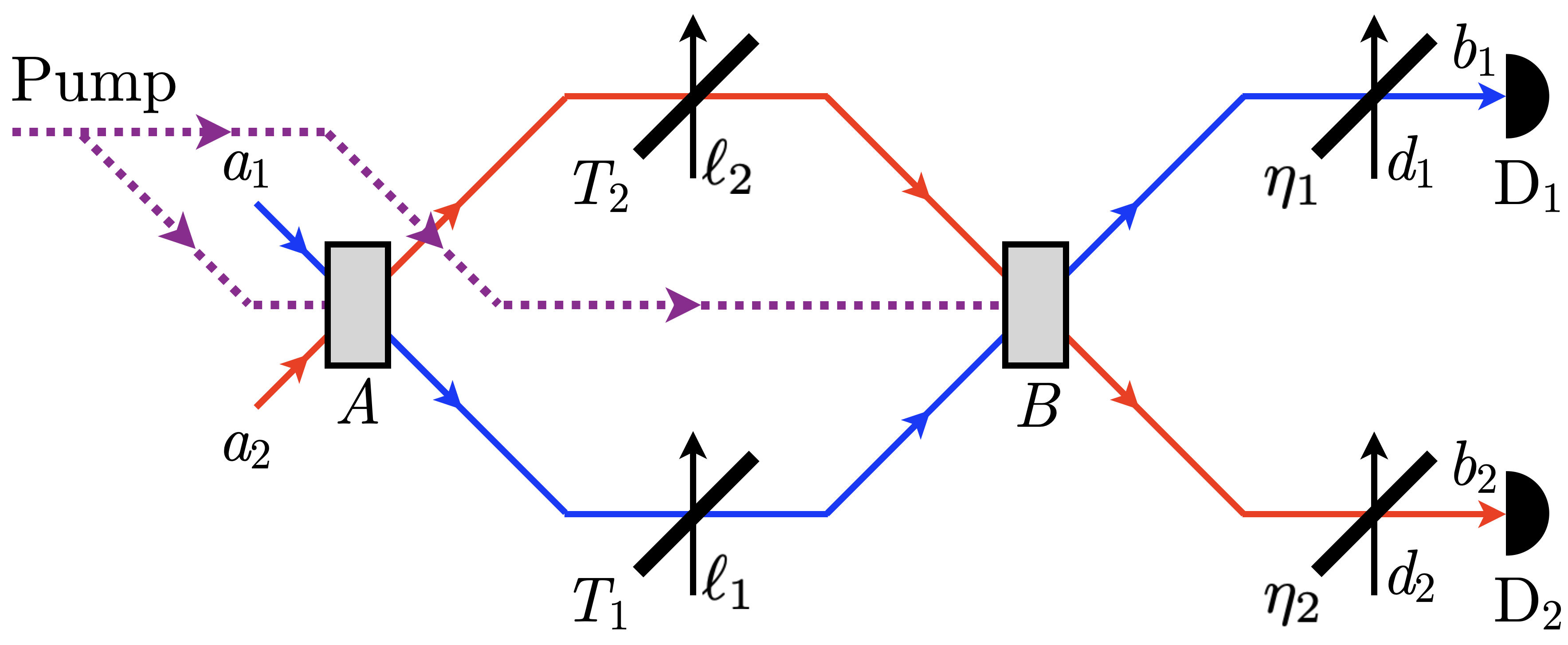}
	\caption{Non-degenerate nonlinear interferometer displaying the input and output modes of each optical process, including the nonlinear interactions, internal losses, and detection inefficiencies. We follow a similar notation as in Ref.~\cite{Giese17}.}
	\label{f:NLI}
\end{figure}

Nonlinear interferometers display a layout similar to their linear counterparts, such as a Mach-Zehnder or Michelson interferometer, except for featuring nonlinear optical media instead of beam splitters~\cite{Chekhova16,Ou20}. These nonlinear media produce correlated photons via two (or more) nonlinear processes in sequence, as illustrated in Fig.~\ref{f:NLI} and further described later in this section. Despite their similar designs, the working principle of these interferometers is entirely different. In the linear case, interference arises from which path light travels inside the interferometer, whereas in the nonlinear case, it arises from which process the correlated photons are produced. Any distinguishability in the correlated photons generated by the two nonlinear processes reduces the interference performance.

To model the single-mode (i.e. single wave vector plus a polarisation direction) non-degenerate nonlinear interferometer shown in Fig.~\ref{f:NLI}, we introduce a matrix describing all physical processes taking place in the interferometer, including the nonlinear interactions, internal losses, and detection inefficiencies. This matrix describes the interferometer output modes $b_1$ and $b_2$ as an evolution of the input modes $a_1$ and $a_2$ in the Heisenberg picture. In Secs.~\ref{s:VisibilityContrast} and~\ref{s:phasesensitivity}, we use this matrix to investigate the nonlinear interferometer performance by introducing seeding in one of the interferometer input modes.

We model the nonlinear processes via optical parametric amplifiers (OPAs), which can be implemented experimentally using parametric down-conversion or four-wave mixing. The action of an OPA whose input modes are $c_1$ and $c_2$ is described by the two-mode squeezing operator $\hat{S}(\xi)=e^{\xi \hat{c}_1^\dagger \hat{c}_2^\dagger-\xi^* \hat{c}_1 \hat{c}_2}$, where $\xi=Ge^{i\varphi}$ is the complex squeezing parameter, $G$ is the real parametric gain, and $\varphi$ is an overall phase~\cite{Olivares12}. Solving the Heisenberg equations of motion yields the Bogoliubov transformation
\begin{equation}
	\begin{bmatrix}
	\hat{c}_{1}'\\
	\hat{c}_{2}'^\dagger
	\end{bmatrix}
	=
	\begin{bmatrix}
	u & v\\
	v^* & u
	\end{bmatrix}
	\begin{bmatrix}
	\hat{c}_1\\
	\hat{c}_2^\dagger
	\end{bmatrix},
	\label{e:OPAA}
\end{equation}
where $c_1'$ and $c_2'$ are the output modes of the OPA, and  $u=\cosh{G}$ and $v=e^{i\varphi}\sinh{G}$ are matrix elements.
The two OPAs~$A$ and~$B$ that make up the nonlinear interferometer in Fig.~\ref{f:NLI} are characterized by squeezing parameters $\xi_j=G_j e^{i\phi_j}$ with $j=A,B$. We then introduce the following real parameters for later convenience
\begin{equation}
    U_j\equiv u_j^2=\cosh^2G_j,\quad  V_j\equiv|v_j|^2=\sinh^2G_j,
\end{equation}
and we note that they satisfy $U_j-V_j=1$. The parameter $V_j$ provides the number of photon pairs generated by an unseeded OPA. The low-gain regime, our regime of interest, occurs whenever $G_j\ll1$.

Internal losses are introduced via beam splitters that mix the input mode $c_1$ with an ancillary mode $c_2$ initialized in the vacuum state. A beam splitter is described by the two-mode mixing operator $\hat{U}(\theta)=e^{\theta(\hat{c}_1^\dagger\hat{c}_2-\hat{c}_2^\dagger\hat{c}_1)}$, where $\theta$ is the beam splitter parameter~\cite{Olivares12}. Note that, when modelling losses, the latter can be taken as real without loss of generality. The evolution of the input modes through the beam splitter yields the following output modes $c_1'$ and $c_2'$,
\begin{equation}
	\begin{bmatrix}
	\hat{c}_1'\\
	\hat{c}_2'
	\end{bmatrix}
	=
	\begin{bmatrix}
	t & r\\
	-r & t
	\end{bmatrix}
	\begin{bmatrix}
	\hat{c}_1\\
	\hat{c}_2
	\end{bmatrix},
	\label{e:BS}
\end{equation}
where $t=\cos{\theta}$ and $r=\sin{\theta}$. We also introduce the transmission $T\equiv t^2$ and reflection $R\equiv r^2$ beam splitter coefficients and note that they satisfy $T+R=1$. The two internal losses in the nonlinear interferometer in Fig.~\ref{f:NLI} are thus characterized by parameters $T_k$ and $R_k$ with $k=1,2$. Finally, detection inefficiencies are also modelled via beam splitters, where the transmission coefficient is given by the detector efficiency $0\leq\eta_k\leq 1$.

Equations~\eqref{e:OPAA} and \eqref{e:BS} allow us to obtain the interferometer output modes $b_1$ and $b_2$ from an overall transformation on the input modes $a_1$, $a_2$ and the ancillary modes $\ell_1$, $\ell_2$, $d_1$, $d_2$ (defined in Fig.~\ref{f:NLI}) in the form
\begin{equation}
	\begin{bmatrix}
	\hat{b}_1\\
	\hat{b}_2^\dagger
	\end{bmatrix}
	=
	\begin{bmatrix}
	A_1 & \alpha_1 & B_1 & \beta_1 & \sqrt{1-\eta_1} & 0\\
	\alpha_2^* & A_2^* & \beta_2^* & B_2^* & 0 & \sqrt{1-\eta_2}\\
	\end{bmatrix}
	\begin{bmatrix}
	\hat{a}_1\\
	\hat{a}_2^\dagger\\
	\hat{\ell}_1\\
	\hat{\ell}_2^\dagger\\	
	\hat{d}_1\\
	\hat{d}_2^\dagger	
	\end{bmatrix},
	\label{e:overallmatrix}
\end{equation}
where we used the following definitions, similar to those introduced in Ref.~\cite{Giese17},
\begin{equation}
\begin{gathered}
	A_1 = \sqrt{\eta_1}\left[\sqrt{T_1U_AU_B}+\sqrt{T_2V_AV_B}e^{-i(\phi_A-\phi_B)}\right],\\
	\alpha_1 = \sqrt{\eta_1}\left[\sqrt{T_1V_AU_B}e^{i(\phi_A-\phi_B)}+\sqrt{T_2U_AV_B}\right]e^{i\phi_B},\\
	B_1 = \sqrt{\eta_1}\sqrt{R_1U_B},\\
	\beta_1 = \sqrt{\eta_1}\sqrt{R_2V_B}e^{i\phi_B},\\
	A_2 = \sqrt{\eta_2}\left[\sqrt{T_1V_AV_B}e^{-i(\phi_A-\phi_B)}+\sqrt{T_2U_AU_B}\right],\\
	\alpha_2 = \sqrt{\eta_2}\left[\sqrt{T_1U_AV_B}+\sqrt{T_2V_AU_B}e^{i(\phi_A-\phi_B)}\right]e^{i\phi_B},\\
	B_2 = \sqrt{\eta_2}\sqrt{R_2U_B},\\
	\beta_2 = \sqrt{\eta_2}\sqrt{R_1V_B}e^{i\phi_B}.
\end{gathered}
\label{e:subsnotsq}
\end{equation}
Equation~\eqref{e:overallmatrix} provides the most general description of a single-mode non-degenerate nonlinear interferometer. In Sec.~\ref{s:VisibilityContrast}, we use this matrix to compute the expected number of output photons, from which we find the interference visibility and contrast.

\section{Visibility and contrast}\label{s:VisibilityContrast}
The interference visibility is probably the most common way to characterize the interferometer performance since it quantifies the appearance of bright and dark fringes. Formally speaking, it is defined as
\begin{equation}
	\mathcal{V}=\frac{\langle \hat{N}_1\rangle^\text{max}-\langle \hat{N}_1\rangle^\text{min}}{\langle \hat{N}_1\rangle^\text{max}+\langle \hat{N}_1\rangle^\text{min}},\label{e:Vis}
\end{equation}
where $\langle\hat{N}_1\rangle^\text{max}$ and $\langle\hat{N}_1\rangle^\text{min}$ are the expected number of output photons at detector D$_1$ when there is constructive and destructive interference, respectively. We can also use detector D$_2$ to define the visibility, but we stick to D$_1$ without loss of generality.

Another figure of merit to characterize the interference performance is the contrast $\mathcal{C}$ given by the difference $\langle \hat{N}_1\rangle^\text{max}-\langle \hat{N}_1\rangle^\text{min}$. This quantity may be useful in experiments with a high detector noise floor, a common scenario when performing imaging with undetected photons~\cite{Lemos14,Cardoso18,Gilaberte21}. Let us calculate $\langle\hat{N}_1\rangle^\text{max}$ and $\langle\hat{N}_1\rangle^\text{min}$ in the simplest case of number-state seeding, and then the resulting visibility and contrast. Later in this section, we repeat the same calculations but in the case of coherent state seeding.

\subsection{Number-state seeding}
Based on the overall transformation matrix in Eq.~\eqref{e:overallmatrix}, we calculate the expected number of photons at detector D$_1$. We first consider a general case where the input modes $a_1$ and $a_2$ are seeded with number states $\ket{n}_{a_1}$ and $\ket{m}_{a_2}$, respectively. The ancillary modes $\ell_1$, $\ell_2$, $d_1$ and $d_2$ are all in the vacuum state.

The expected number of photons $\langle \hat{N}_1\rangle$ calculated from the inner product $\bra{n}_{a_1}\bra{m}_{a_2}\hat{b}_1^\dagger\hat{b}_1\ket{n}_{a_1}\ket{m}_{a_2}$ is equal to (see Appendix~\ref{a:Neta1number})
\begin{align}
    \langle\hat{N}_1\rangle=&\ n|A_1|^2+|\alpha_1|^2(m+1)+|\beta_1|^2\notag\\
    =&\ \eta_1 n \Big(T_1U_AU_B+T_2V_AV_B\notag\\
    &\phantom{n\eta_1\Big(}+2\sqrt{T_1T_2U_AU_BV_AV_B}\cos\phi\Big)\notag\\ 
	&+\eta_1(m+1)\Big(T_1V_AU_B+T_2U_AV_B\notag\\
	&\phantom{+\eta_1(m+1)\Big(}+2\sqrt{T_1T_2U_AU_BV_AV_B}\cos\phi\Big)\notag\\
	&+\eta_1R_2V_B,\label{e:Neta1}
\end{align}
where $A_1$, $\alpha_1$ and $\beta_1$ are given in Eq.~\ref{e:subsnotsq} and the interferometer relative phase is
\[
	\phi=\phi_A-\phi_B.
\]
Note that $\langle \hat{N}_1\rangle$ only depends on the phase difference between the OPAs $A$ and $B$, and not the individual values of $\phi_A$ and $\phi_B$.

If both $n$ and $m$ vanish, we recover the expression for $\langle \hat{N}_1\rangle$ reported in Ref.~\cite{Giese17}, Eq.~(14), for an unseeded nonlinear interferometer. If $n$ vanishes and $\eta_1$ equals unity, we arrive at
\begin{align*}
	\langle \hat{N}_{1,\eta_1=1}\rangle=&\ \frac{1}{2}(m+1)\sinh (2G_A)\sinh (2G_B)\sqrt{T_1T_2}\cos\phi\\
	&+(m+1)\sinh^2(G_B)\cosh^2(G_A)T_2\\
	&+(m+1)\sinh^2(G_A)\cosh^2(G_B)T_1\\
	&+\sinh^2(G_B)R_2,
\end{align*}
which is the expected number of signal photons reported in Ref.~\cite{Michael21}, Eq.~(1). Equation~\eqref{e:Neta1} provides the most general expression of the expected number of photons at detector D$_1$ when seeding with number states because it takes into account seeding both $a_1$ and $a_2$ input modes, internal losses, inefficient detection, and parametric gain unbalancing.

Based on Eq.~\eqref{e:Neta1}, $\langle\hat{N}_1\rangle^\text{max}$ and $\langle\hat{N}_1\rangle^\text{min}$ are found at $\phi=0$ and $\phi=\pi$, respectively. From Eq.~\eqref{e:Vis}, the visibility then reads
\begin{align}
	\mathcal{V}=&\ 2(n+m+1)\sqrt{T_1T_2U_AU_BV_AV_B}\notag\\ 
	&\Big/\Big[n\left(T_1U_AU_B+T_2V_AV_B\right)\notag\\
	&\phantom{\Big/\Big[}+(m+1)\left(T_1V_AU_B+T_2U_AV_B\right)+R_2V_B\Big].\label{e:Vnumber}
\end{align}
At first sight, the contributions from $n$ and $m$ to Eq.~\eqref{e:Vnumber} seem to have comparable effects. However, in the low-gain regime, where $G_j\ll1$ (or equivalently $V_j\ll1$ and $U_j\approx1$), 
all the terms in the denominator of Eq.~\eqref{e:Vnumber} have a negligible weight compared with $nT_1U_AU_B\approx nT_1$. Therefore, if $n\ne0$, we get a large denominator in Eq.~\eqref{e:Vnumber}, leading to a vanishing visibility. Another way to understand the effect of seeding mode $a_1$ with $n$ photons is by noting that these $n$ non-interfering photons end up illuminating detector D$_1$, which increases both $\langle\hat{N}_1\rangle^\text{max}$ and $\langle\hat{N}_1\rangle^\text{min}$ and deteriorates the visibility. We conclude that seeding mode $a_1$ is detrimental in the low-gain regime for visibility-based applications, and focus only on seeding the undetected mode $a_2$ by setting $n=0$ in the following calculations.

Interestingly, Eq.~\eqref{e:Vnumber} (with $n=0$) can be rewritten as
\begin{equation}
    \mathcal{V}=\cfrac{2\gamma}{1+\gamma^2+\cfrac{R_2}{(m+1)T_2U_A}},\label{e:Vnumberdelta}
\end{equation}
where we defined the parameter $\gamma$ as
\begin{equation}
    \gamma=\sqrt{\frac{T_1V_AU_B}{T_2U_AV_B}}.
\end{equation}
According to Eq.~\eqref{e:Vnumberdelta}, seeding the undetected mode with $m$ photons can mitigate the losses in this mode, quantified via the $R_2/T_2$ ratio in the denominator, which leads to an increase in the visibility. In the ideal case of a lossless undetected mode, i.e. $R_2/T_2=0$, seeding does not affect the visibility. Moreover, these $m$ seeding photons cannot compensate for losses in the detected mode $a_1$, as can be anticipated intuitively.

The parameter $\gamma$ can be interpreted as a balance parameter. For example, in the low-gain regime, it reduces to $\gamma\approx\sqrt{T_1V_A/(T_2V_B)}$, which is the square root of the transmission ratio between the detected and undetected modes, $T_1/T_2$, times the generated photon pair ratio between the OPAs $A$ and $B$, $V_A/V_B$. Moreover, if $T_1\approx T_2$ and $V_A\approx V_B$, the visibility reduces to
\[
    \mathcal{V}\approx\cfrac{2}{2+\cfrac{R_2}{(m+1)T_2}},
\]
which is a simple and useful expression to estimate the visibility of a nonlinear interferometer in the low-gain regime, including losses and seeding in the undetected mode. 

Regarding the contrast, from Eq.~\eqref{e:Neta1}, $\mathcal{C}$ reduces to
\begin{equation}
	\mathcal{C}=4{\eta_1}(n+m+1)\sqrt{T_1T_2U_AU_BV_AV_B},\label{e:Cnumber}
\end{equation}
where we can make $n=0$ without loss of generality since $n$ and $m$ have the same effect on $\mathcal{C}$, and therefore the contribution from $n$ can be counted within $m$ if needed.

\begin{figure}[b!]
	\centering
	\includegraphics[width=0.5\textwidth]{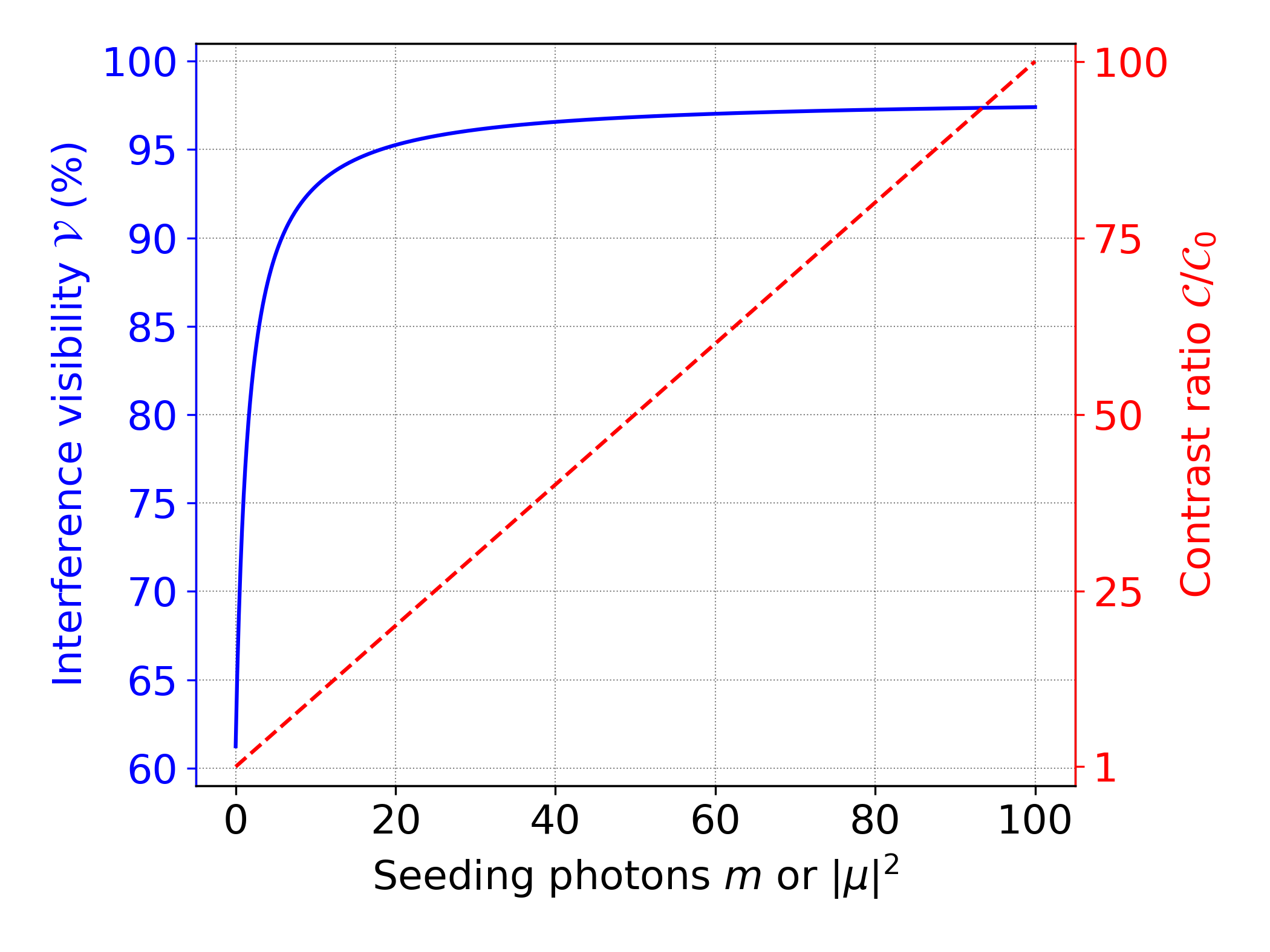}
	\caption{Interference visibility (solid blue) and contrast ratio (dashed red) as a function of the seeding photon number $m$ or $|\mu|^2$, depending on the seeding state in mode $a_2$, a number state $\ket{m}_{a_2}$ or a coherent state $\ket{\mu}_{a_2}$, respectively. Mode $a_1$ is in the vacuum state in both cases. $\mathcal{C}_0$ is the contrast without seeding ($m=0$).}
	\label{f:VisibilityContrast}
\end{figure}

Figure~\ref{f:VisibilityContrast} shows the visibility and contrast in Eqs.~\eqref{e:Vnumber} and~\eqref{e:Cnumber}, respectively, as a function of the seeding photon number $m$. The contrast is illustrated as a ratio relative to the contrast without seeding ($m=0$), $\mathcal{C}_0$. We use typical parameters for internal losses ($T_1=0.6$, $T_2=0.4$), detection efficiency ($\eta_1=0.3$) and parametric gains ($G_A=G_B=10^{-3}$). These parameters represent the substantial internal losses and highly inefficient detectors that one would expect on average in a realistic experimental setup~\cite{Gemmell22}. Moreover, we consider parametric gains in the low-gain regime, although our expressions~\eqref{e:Vnumber} and~\eqref{e:Cnumber} are fully valid for any value of $G_A$ and $G_B$, as long as the pump is not depleted~\cite{Florez18}.

We see in Fig.~\ref{f:VisibilityContrast} that both visibility and contrast increase as a function of $m$. The visibility rapidly approaches 100\%, reaching a maximum of 97.4\%. Even a few tens of seeding photons are enough to improve the visibility from 61\% to $>$95\%, which represents an important enhancement for visibility-based applications. A 100\% visibility is not reached via seeding because of the asymmetry between $T_1$ and $T_2$. In contrast, if the losses are symmetric, e.g. $T_1=T_2=0.4$, the visibility is 99.3\% for $m=100$ (see Appendix~\ref{a:VT1T2}), and will continue to increase asymptotically towards 100\% with higher seeding photon number. As for the contrast, it improves by a factor equal to $m+1$. These results prove that seeding has a remarkable potential to enhance the overall interference performance.

\subsection{Coherent state seeding}
So far, we have focused on seeding mode $a_2$ with a number state $\ket{m}_{a_2}$. Despite the resulting simplicity in the visibility and contrast derivations, generating number states on-demand is challenging in practice, making our theoretical results difficult to implement in experimental terms. Therefore, we now investigate a nonlinear interferometer seeded by a coherent state $\ket{\mu}_{a_2}$ in mode $a_2$. An attenuated laser can generate this state, making it simple to incorporate into a nonlinear interferometer setup. We dismiss seeding mode $a_1$ based on similar arguments as in the number-state seeding case. However, we found that seeding the detected mode with a Gaussian state may increase the visibility if the seeding offsets the loss in the interferometer. This scenario can be concisely described via the Gaussian formalism~\cite{Olivares12,Vallone19},  but the resulting expressions are rather complicated, as reported elsewhere~\cite{Plick10}, and beyond the scope of this paper.

We start by calculating the expected number of photons at detector D$_1$, and then the visibility and contrast. The expression for $\langle\hat{N}_1\rangle$ has the same form of Eq.~\eqref{e:Neta1} but with $m$ substituted by the mean photon number $|\mu|^2$ after setting $n=0$ (see Appendix~\ref{a:Neta1coherent}). Therefore, the visibility and contrast are dictated by Eqs.~\eqref{e:Vnumber} and~\eqref{e:Cnumber} in the case of coherent seeding after applying this substitution. Likewise, the results in Fig.~\ref{f:VisibilityContrast} and the conclusions that we drew for number-state seeding are equally applicable to coherent state seeding.

Note that the visibility and contrast depend only on the expected number of output photons and are not influenced by higher-order moments. Hence, the photon number fluctuations that differentiate number from coherent states do not affect the interference performance. In Sec.~\ref{s:phasesensitivity}, we shall see that this equivalence is still valid in the low-gain regime when dealing with the phase sensitivity and SNR of a seeded nonlinear interferometer.

\section{Phase sensitivity and signal-to-noise ratio}\label{s:phasesensitivity}
Besides the interference performance quantified via the visibility and contrast, another interesting property of nonlinear interferometers is their sensitivity to detect phase shifts. This phase sensitivity can be quantified using the phase variance $\Delta\phi^2$ and the number of photons at detector D$_1$ as~\cite{Gerry}
\begin{equation}
	\Delta\phi^2=\frac{\Delta\hat{N}_1^2}{(\partial\langle\hat{N}_1\rangle/\partial\phi)^2}\label{e:Deltaphisq},
\end{equation}
where $\Delta\hat{N}_1^2$ is the variance of $\hat{N}_1$ computed as $\langle\hat{N}_1^2\rangle-\langle\hat{N}_1\rangle^2$. Another method to calculate $\Delta\phi^2$ is using estimation theory, where the Fisher information provides the best phase sensitivity achievable when we infer the value of $\phi$ from measurements of the observable $\hat{N}_1$~\cite{Giovannetti11,Gabbrielli15,Giese17,Florez18}. However, we stick to the widely-used error propagation formula in Eq.~\eqref{e:Deltaphisq} for comparison reasons~\cite{Yurke86,Marino12,Manceau17,Szigeti17,Michael21}.

The last property of nonlinear interferometers that we shall investigate is the SNR, which quantifies the statistical fluctuations in the expected number of photons due to the quantum nature of light. A higher SNR means fewer fluctuations in the detector readings, which leads to more precise results. From the expected number of photons $\langle\hat{N}_1\rangle$ and its variance $\Delta\hat{N}_1^2$, the SNR is given by
\begin{equation}
    \mathrm{SNR}=\frac{\langle\hat{N}_1\rangle^2}{\Delta\hat{N}_1^2}.\label{e:SNRdef}
\end{equation}
Let us derive phase sensitivity and SNR expressions in the case of number and coherent state seeding.

\subsection{Number-state seeding}
By focusing on seeding mode $a_2$ with a number state $\ket{m}_{a_2}$, we find the variance of $\hat{N}_1$ with the help of the overall transformation matrix in Eq.~\eqref{e:overallmatrix} and the derivative of $\langle\hat{N}_1\rangle$ with respect to $\phi$. We then substitute these expressions into Eq.~\eqref{e:Deltaphisq} and obtain $\Delta\phi^2$ (see Appendix~\ref{a:phasesensitivity_number}). The resulting expression for the phase sensitivity is
\begin{align}
    \Delta\phi^2=&\ \Big[|\alpha_1|^2(m+1)\left(|A_1|^2+|B_1|^2+1-\eta_1\right)\notag\\
    &\phantom{\Big[}+|\beta_1|^2\left(|A_1|^2+m|\alpha_1|^2+|B_1|^2+1-\eta_1\right)\Big]\notag\\
    &\phantom{\Big[}\Big/ \Big[4\eta_1^2(m+1)^2T_1T_2U_AU_BV_AV_B\sin^2\phi\Big].\label{e:Deltaphisq_number}
\end{align}
Please note that the coefficients $|A_1|^2$ and $|\alpha_1|^2$ explicitly depend on $\phi$ (see Appendix~\ref{a:Neta1number}). Also note that, similar to $\langle \hat{N}_1\rangle$, $\Delta\phi^2$ only depends on the difference between $\phi_A$ and $\phi_B$ and not the individual values of these two phases.

Equation~\eqref{e:Deltaphisq_number} displays at least one minimum as a function of $\phi$. For a lossless ($T_1=T_2=1$), detection efficient ($\eta_1=1$) and gain balanced ($G_A=G_B=G$) nonlinear interferometer, such a minimum is located at $\phi=\pi$ and is equal to
\begin{equation}
	\Delta\phi^2_\text{QL}=\frac{1}{4(m+1)\sinh^2G(\sinh^2G+1)}.\label{e:DeltaphisqQL_number}
\end{equation}
We identify $\Delta\phi^2_\text{QL}$ as the maximum phase sensitivity that a number-state seeded nonlinear interferometer can achieve according to quantum theory and the error propagation formula in Eq.~\eqref{e:Deltaphisq}. Therefore, $\Delta\phi^2_\text{QL}$ is also known as the Heisenberg or quantum-limited (QL) phase sensitivity. 

\begin{figure}[t!]
	\centering
	\includegraphics[width=0.5\textwidth]{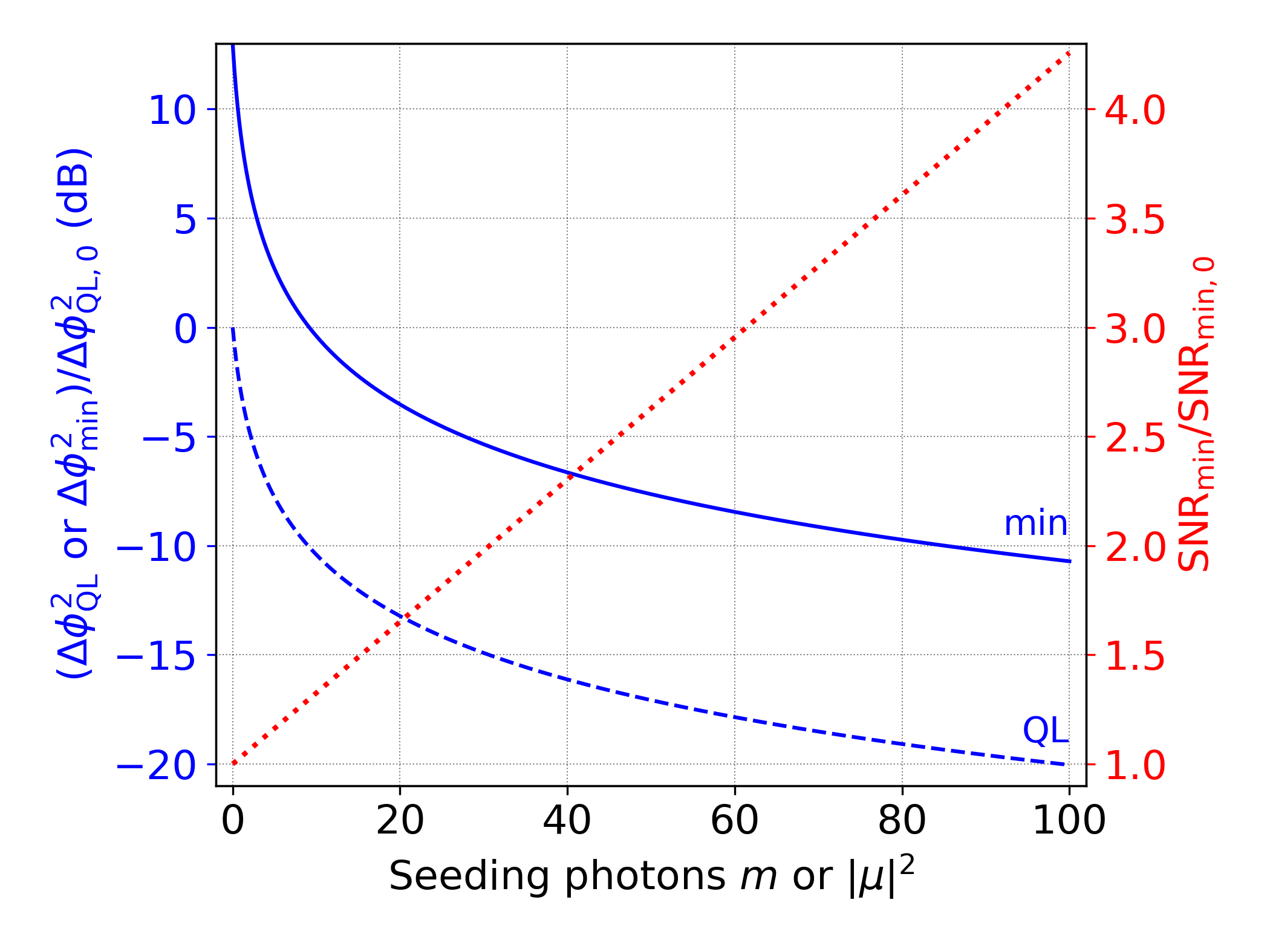}
	\caption{Maximum phase sensitivity $\Delta\phi^2_\text{min}$ (solid blue) of the nonlinear interferometer in Fig.~\ref{f:NLI} when mode $a_2$ is seeded by a number or a coherent state with $m$ or $|\mu|^2$ mean photon number, respectively. We also present the quantum-limited phase sensitivity $\Delta\phi^2_\text{QL}$ (dashed blue) for a seeded nonlinear interferometer. Both $\Delta\phi^2_\text{min}$ and $\Delta\phi^2_\text{QL}$ are normalized to the quantum-limited phase sensitivity of an unseeded ($m=0$) nonlinear interferometer, $\Delta\phi^2_{\text{QL},0}$. Finally, we present the minimum signal-to-noise ratio SNR$_\mathrm{min}$ (dotted red) normalized to the unseeded SNR, SNR$_{\mathrm{min},0}$. Mode $a_1$ is in the vacuum state in all cases.}
	\label{f:DeltaphiSqmin}
\end{figure}

For arbitrary values of $T_k$, $\eta_1$, and $G_j$, $\Delta\phi^2$ displays two identical minima $\Delta\phi^2_\text{min}$ symmetrically located with respect to $\phi=\pi$. We numerically find one of these minima and study the effect of seeding when there are internal losses, detection inefficiencies, and balanced parametric gains. The results are shown in Fig.~\ref{f:DeltaphiSqmin} as a function of $m$ using the same values of $T_k$, $\eta_1$ and $G_j$ as in Fig.~\ref{f:VisibilityContrast}. For comparison reasons, we normalize $\Delta\phi^2_\text{min}$ and $\Delta\phi^2_\text{QL}$ with respect to the quantum-limited phase sensitivity of an unseeded nonlinear interferometer, $\Delta\phi^2_{\text{QL},0}$.

In Fig.~\ref{f:DeltaphiSqmin}, we observe that $\Delta\phi^2_\text{min}/\Delta\phi^2_{\text{QL},0}$ decreases as a function of $m$, starting above 10 dB for $m=0$ and reaching values below $-10$~dB for $m=100$. Interestingly, $\Delta\phi^2_\text{min}/\Delta\phi^2_{\text{QL},0}$ is below 0~dB starting at $m\sim10$. This result is remarkable because it implies that a seeded nonlinear interferometer can outperform the quantum-limited phase sensitivity of an unseeded one with a few tens of seeding photons, even in the presence of substantial internal losses and highly inefficient detectors.

Another observation in Fig.~\ref{f:DeltaphiSqmin} is the fact that $\Delta\phi^2_\text{min}$ is always greater than $\Delta\phi^2_\text{QL}$, as expected from the imperfections in the nonlinear interferometer. However, as previously mentioned, we can compensate for these imperfections by further increasing $m$. For example, $\Delta\phi^2_\text{QL}/\Delta\phi^2_{\text{QL},0}$ is approximately $-10$~dB for $m\sim10$, but we can reach the same phase sensitivity with a lossy and detection inefficient nonlinear interferometer by seeding with $m\sim100$ photons. Therefore, quantum-limited phase sensitivities can in principle be achieved with an imperfect nonlinear interferometer, assuming we have arbitrary seeding photon numbers at our disposal.

Regarding the SNR in Eq.~\eqref{e:SNRdef}, and from the expected number of photons $\langle\hat{N}_1\rangle$ calculated in Sec.~\ref{s:VisibilityContrast}, Eq.~\eqref{e:Neta1}, and its variance $\Delta\hat{N}_1^2$ [see Appendix~\ref{a:phasesensitivity_number}, Eq.~\eqref{e:VarNeta1}], we obtain
\begin{align}
    \mathrm{SNR}=&\ \Big[|\alpha_1|^2(m+1)+|\beta_1|^2\Big]^2\notag\\
    &\phantom{\Big[}\Big/\Big[|\alpha_1|^2(m+1)\left(|A_1|^2+|B_1|^2+1-\eta_1\right)\notag\\
    &\phantom{\Big[\Big/}+|\beta_1|^2\left(|A_1|^2+m|\alpha_1|^2+|B_1|^2+1-\eta_1\right)\Big].\label{e:SNR_number}
\end{align}
As in the case of $\langle\hat{N}_1\rangle$ and $\Delta\phi^2$ in Eqs.~\eqref{e:Neta1} and~\eqref{e:Deltaphisq_number}, respectively, the SNR depends on the relative phase $\phi$. In particular, Eq.~\eqref{e:SNR_number} exhibits a minimum at $\phi=\pi$ and maxima at $\phi=0$ and $2\pi$. The minimum SNR (SNR$_\mathrm{min}$) is normalized to the unseeded minimum SNR (SNR$_{\mathrm{min},0}$), and shown as a function of $m$ in Fig.~\ref{f:DeltaphiSqmin}. The same values of $T_k$, $\eta_1$ and $G_j$ are used as in Fig.~\ref{f:VisibilityContrast}. According to Fig.~\ref{f:DeltaphiSqmin}, SNR$_\mathrm{min}$ increases linearly with the number of seeding photons, with around a four-fold improvement at $m\sim100$ photons compared to the unseeded case. Therefore, we must expect fewer statistical fluctuations in the expected number of photons at detector D$_1$ when we seed the nonlinear interferometer.

\subsection{Coherent state seeding}
As discussed in Sec.~\ref{s:VisibilityContrast}, producing an arbitrary number state is extremely challenging from an experimental point of view, so we once again focus on seeding mode $a_2$ with a coherent state $\ket{\mu}_{a_2}$. The phase sensitivity for a nonlinear interferometer seeded by $\ket{\mu}_{a_2}$ is given by (see Appendix~\ref{a:phasesensitivity_coherent})
\begin{align}
    \Delta\phi^2=&\ \Big[|\mu|^2|\alpha_1|^4\notag\\
    &\phantom{\Big[}+|\alpha_1|^2\left(|\mu|^2+1\right)\left(|A_1|^2+|B_1|^2+1-\eta_1\right)\notag\\
    &\phantom{\Big[}+|\beta_1|^2\left(|A_1|^2+|\mu|^2|\alpha_1|^2+|B_1|^2+1-\eta_1\right)\Big]\notag\\
    &\phantom{\Big[}\Big/ \Big[4\left(|\mu|^2+1\right)^2\eta_1^2T_1T_2U_AU_BV_AV_B\sin^2\phi\Big].\label{e:Deltaphisq_coherent}
\end{align}
For an ideal interferometer, $\Delta\phi^2_\text{QL}$ is given in Eq.~\eqref{e:DeltaphisqQL_number} after substituting $m$ by $|\mu|^2$, while for a non-ideal interferometer $\Delta\phi^2_\text{min}$ is found in a similar way to the number-state seeding case. The results for both $\Delta\phi^2_\text{QL}$ and $\Delta\phi^2_\text{min}$ as a function of $|\mu|^2$ for the same values of $T_k$, $\eta_1$ and $G_j$ as in Fig.~\ref{f:VisibilityContrast} overlap with the ones already presented in Fig.~\ref{f:DeltaphiSqmin} for number-state seeding.

To understand this overlapping we need to take a look at the variance of $\hat{N}_1$. When seeding with a coherent state, $\Delta\hat{N}_1^2$ displays the extra term $|\mu|^2|\alpha_1|^4$ compared with number-state seeding [see Appendix~\ref{a:phasesensitivity_coherent}, Eq.~\eqref{e:VarNeta1}]. The reason for this discrepancy between seeding with a number versus a coherent state is that coherent states add extra noise to the number of photons generated by an OPA. This extra noise shows up in $\Delta\hat{N}_1^2$, but not in $\langle\hat{N}_1\rangle$. However, the weight of this extra noise in the low-gain regime is negligible because $|\alpha_1|^4$ contains terms of the form $V_j^2\ll1$, leading to $|\alpha_1|^4\ll1$. As a result, we obtain the same photon number variance, and therefore the same phase sensitivity, as in the case of number-state seeding. Thus, the observations made for the phase sensitivities of an ideal and non-ideal number-state seeded nonlinear interferometer are valid in the case of coherent state seeding.

For the SNR we expect from $\langle\hat{N}_1\rangle$ [see Appendix~\ref{a:Neta1coherent}, Eq.~\eqref{e:Neta1_coherent}], and $\Delta\hat{N}_1^2$ [see Appendix~\ref{a:phasesensitivity_coherent}, Eq.~\eqref{e:VarNeta1_coherent}], the following expression according to its definition in Eq.~\eqref{e:SNRdef},
\begin{align}
    \mathrm{SNR}=&\ \Big[|\alpha_1|^2(|\mu|^2+1)+|\beta_1|^2\Big]^2\notag\\
    &\phantom{\Big[}\Big/\Big[|\mu|^2|\alpha_1|^4\notag\\
    &\phantom{\Big[\Big/}+|\alpha_1|^2(|\mu|^2+1)\left(|A_1|^2+|B_1|^2+1-\eta_1\right)\notag\\
    &\phantom{\Big[\Big/}+|\beta_1|^2\left(|A_1|^2+|\mu|^2|\alpha_1|^2+|B_1|^2+1-\eta_1\right)\Big].\label{e:SNR_coherent}
\end{align}
Once again we get the extra term $|\mu|^2|\alpha_1|^4$ in the denominator of Eq.~\eqref{e:SNR_coherent}, which vanishes in the low-gain regime as previously discussed. Therefore, the SNRs in the cases of number and coherent state seeding overlap with one another, and the observations introduced earlier for the SNR based on Fig.~\ref{f:DeltaphiSqmin} apply to coherent state seeding as well.

\section{Conclusions}\label{s:conclusions}
We presented a comprehensive study on how seeding can significantly improve the performance of a lossy and detection inefficient nonlinear interferometer. Unlike similar works, we focused on the effect that a varying seeding photon number has on the visibility, contrast, phase sensitivity, and SNR. Based on an overall transformation matrix describing the nonlinear interferometer, and considering number and coherent seeding states, we provided analytical expressions for all of these interferometer properties. We found that by seeding the undetected input mode with a few tens of photons, the visibility and phase sensitivity are significantly enhanced in the presence of substantial internal losses and highly inefficient detectors. Likewise, the contrast and SNR displayed a linear increase with the number of seeding photons, reaching a multiple-fold improvement compared to the unseeded case. Interestingly, we observed that the enhancement in these four properties is the same in the low-gain regime regardless of whether the nonlinear interferometer is seeded by a number or a coherent state, as long as they both have the same mean photon number. This means that our theoretical results can be implemented in the laboratory using an attenuated laser as the seeding source, for instance, avoiding the use of number seeding states at all. 

Enhancing nonlinear interferometer performance via seeding the undetected photons opens up new opportunities in quantum imaging, metrology, and spectroscopy. Although the requirement of seeding the undetected wavelength might present difficulties in the mid-IR, detection may still take place in the visible or near-IR, making this scheme suitable for applications such as infrared imaging and spectroscopy. These results pave the way to achieve quantum-enhanced metrology with current experimental realizations of nonlinear interferometers in the presence of internal losses and inefficient detectors.

\appendix
\begin{widetext}
\section{Expected number of output photons when seeding with number states}\label{a:Neta1number}
\noindent The expected number of photons $\langle\hat{N}_1\rangle$ at detector D$_1$ is given by
\[
	\langle\hat{N}_1\rangle=\bra{n}_{a_1}\bra{m}_{a_2}\hat{b}_1^\dagger\hat{b}_1\ket{n}_{a_1}\ket{m}_{a_2}.
\]
To obtain the result in Eq.~\eqref{e:Neta1}, we exploit the overall transformation matrix in Eq.~\eqref{e:overallmatrix} to calculate the action of $\hat{b}_1$ on the initial state $\ket{n}_{a_1}\ket{m}_{a_2}$,
\begin{equation}
	\hat{b}_1\ket{n}_{a_1}\ket{m}_{a_2}=A_1\sqrt{n}\ket{n-1}_{a_1}\ket{m}_{a_2}+\alpha_1\sqrt{m+1}\ket{n}_{a_1}\ket{m+1}_{a_2}+\beta_1\ket{n}_{a_1}\ket{m}_{a_2}\ket{1}_{\ell_2}.\label{e:bnm}
\end{equation}
We write explicitly only those quantum states different from vacuum to simplify our notation. For example, the initial state $\ket{n}_{a_1}\ket{m}_{a_2}$ is indeed $\ket{n}_{a_1}\ket{m}_{a_2}\ket{0}_{\ell_1}\ket{0}_{\ell_2}\ket{0}_{d_1}\ket{0}_{d_2}$. We adopt this convention in all derivations throughout this paper. 

Now, we take the inner product between $\bra{n}_{a_1}\bra{m}_{a_2}\hat{b}_1^\dagger$ and $\hat{b}_1\ket{n}_{a_1}\ket{m}_{a_2}$ based on Eq.~\eqref{e:bnm} to find
\begin{equation}
	\bra{n}_{a_1}\bra{m}_{a_2}\hat{b}_1^\dagger\hat{b}_1\ket{n}_{a_1}\ket{m}_{a_2}=n|A_1|^2+|\alpha_1|^2(m+1)+|\beta_1|^2\label{e:nmbbnm}
\end{equation}
where
\begin{gather}
	|A_1|^2=\eta_1\left(T_1U_AU_B+T_2V_AV_B+2\sqrt{T_1T_2U_AU_BV_AV_B}\cos\phi\right),\label{e:A1sq}\\
	|\alpha_1|^2=\eta_1\left(T_1V_AU_B+T_2U_AV_B+2\sqrt{T_1T_2U_AU_BV_AV_B}\cos\phi\right),\label{e:alpha1sq}\\
	|\beta_1|^2=\eta_1R_2V_B.\label{e:beta1sq}
\end{gather}
For future reference,
\[
	|B_1|^2=\eta_1R_1U_B.
\]
Finally, we substitute Eqs.~\eqref{e:A1sq} through~\eqref{e:beta1sq} into Eq.~\eqref{e:nmbbnm} to get $\langle\hat{N}_1\rangle$ in Eq.~\eqref{e:Neta1}.

\section{Visibility for varying internal losses}\label{a:VT1T2}
\noindent If we assume that the losses inside the interferometer are symmetric, i.e. $T_1=T_2$, the interference visibility closely approaches 100\% as a function of the seeding photons for increasing values of $T_1=T_2$. We illustrate this scenario in Fig.~\ref{f:VT1T2}(a) via a contour plot accompanied by cross sections at four different $T_1=T_2$ values.

If we fix $T_1=0.6$ and vary $T_2$, we observe in Fig.~\ref{f:VT1T2}(b) that the maximum visibility is obtained for $T_2$ around~0.6. Other values of $T_2$ lead to lower visibilities, although they may lead to faster increasing visibility as a function of the seeding photons, like $T_2=0.8$ with $m$ or $|\mu|^2$ up to 10 photons. Finally, if we fix $T_2=0.4$ and vary $T_1$, as we show in Fig.~\ref{f:VT1T2}(c), we observe again the maximum visibility for $T_1$ around 0.4. Since having the same transmissions $T_1$ and $T_2$ is unlikely in an experiment, we focus on asymmetric internal losses in the main text.
\begin{figure}[b!]
	\centering
	\includegraphics[width=0.3297\textwidth]{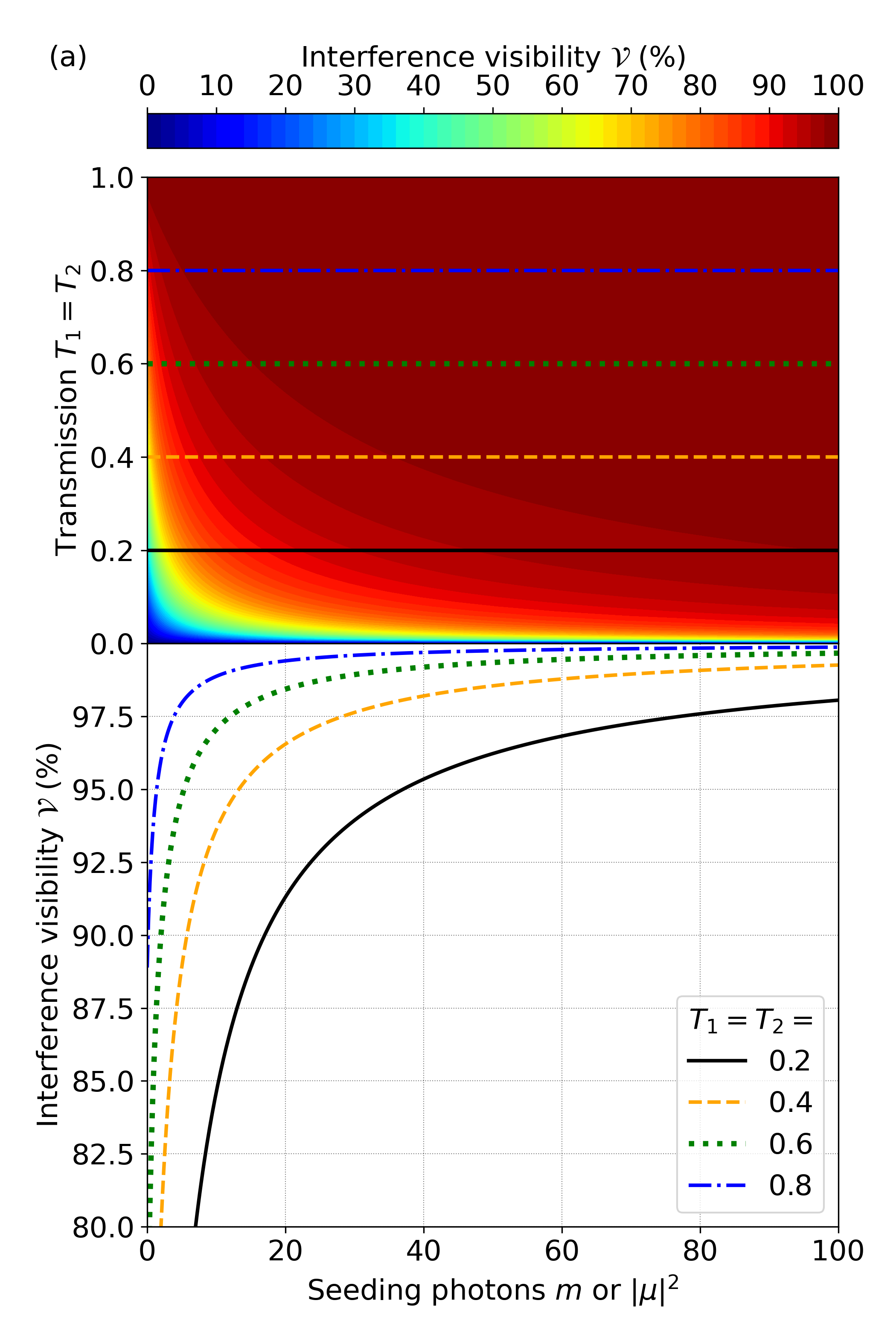}
	\includegraphics[width=0.3297\textwidth]{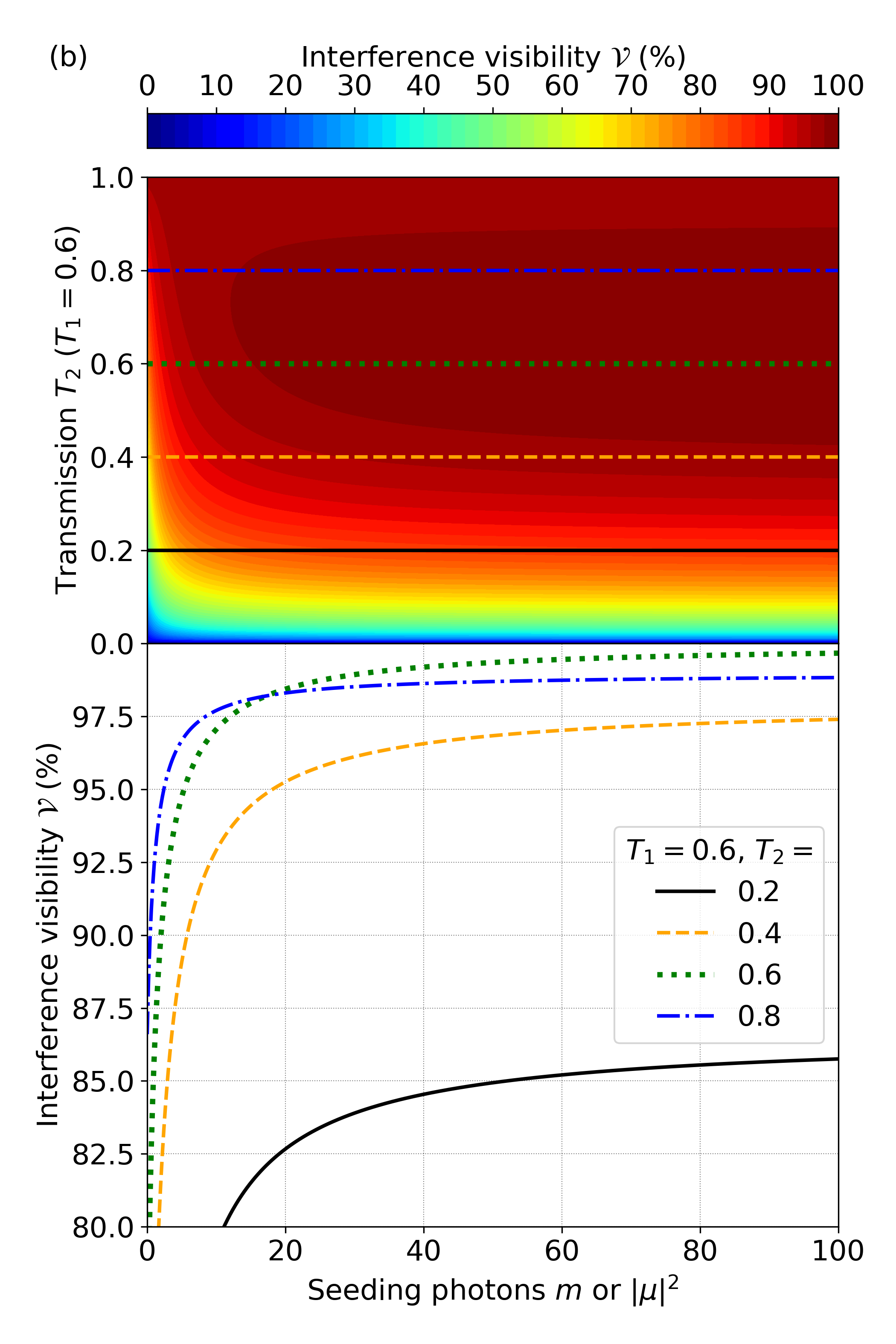}
	\includegraphics[width=0.3297\textwidth]{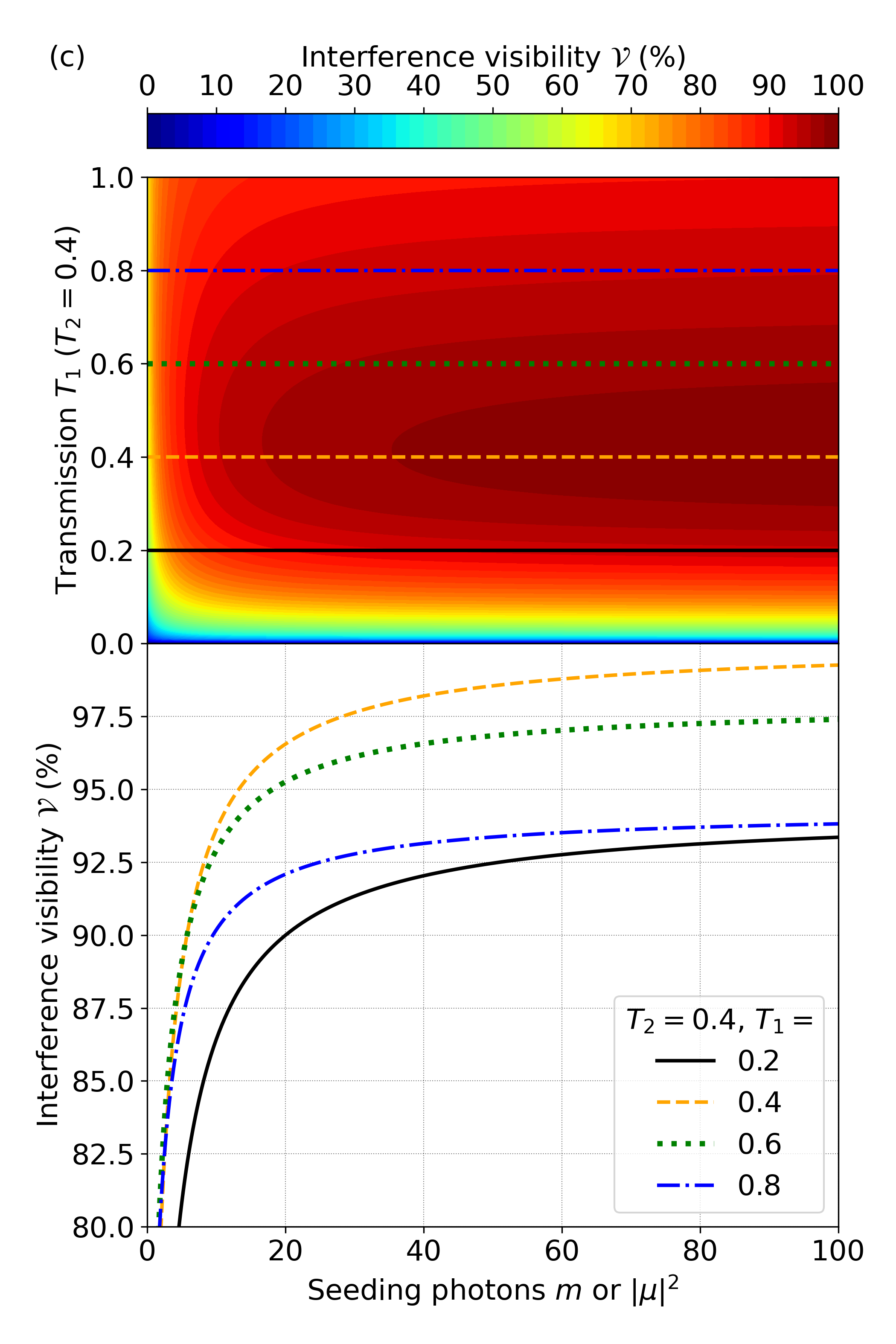}
	\caption{Visibility for varying values of $T_1$ and $T_2$ as a function of the seeding photons. The top panels display contour visibility plots as a function of the seeding photons and the transmission $T_1=T_2$ (a), $T_2$ (b), and $T_1$ (c). The bottom panels show cross sections of the contour plots at the indicated transmission values in the top panels using the same line colours and styles.}
	\label{f:VT1T2}
\end{figure}

\section{Expected number of output photons when seeding with a coherent state}\label{a:Neta1coherent}
\noindent We assume that mode $a_2$ in Fig.~\ref{f:NLI} is initially in the coherent state $\ket{\mu}_{a_2}$, while mode $a_1$ is in vacuum. Let us obtain the expected number of photons at detector D$_1$. By applying $\hat{b}_1$ on the initial state $\ket{\mu}_{a_2}$, and exploiting the overall transformation matrix in Eq.~\eqref{e:overallmatrix}, we obtain \begin{equation*}
	\hat{b}_1\ket{\mu}_{a_2}=\alpha_1e^{-\frac{|\mu|^2}{2}}\sum_{q=0}^\infty\frac{\mu^q}{\sqrt{q!}}\sqrt{q+1}\ket{q+1}_{a_2}+\beta_1\ket{\mu}_{a_2}\ket{1}_{\ell_2}.
\end{equation*}
Recall that we are using the convention of writing only those quantum states that are different from vacuum. The expected number of photons at detector D$_1$ is then
\[
	\langle \hat{N}_1\rangle=\bra{\mu}_{a_2}\hat{b}_1^\dagger\hat{b}_1\ket{\mu}_{a_2}=|\alpha_1|^2e^{-|\mu|^2}\sum_{q=0}^\infty\frac{|\mu|^{2q}}{q!}(q+1)+|\beta_1|^2,
\]
which reduces to
\begin{equation}
	\langle \hat{N}_1\rangle=|\alpha_1|^2(|\mu|^2+1)+|\beta_1|^2.\label{e:Neta1_coherent}
\end{equation}
By using Eqs.~\eqref{e:alpha1sq} and~\eqref{e:beta1sq}, we finally arrive to
\begin{equation}
	\langle\hat{N}_1\rangle=\eta_1(|\mu|^2+1)\left(T_1V_AU_B+T_2U_AV_B+2\sqrt{T_1T_2U_AU_BV_AV_B}\cos\phi\right)+\eta_1R_2V_B.\label{e:Neta1coh}
\end{equation}
We obtained the same expression for $\langle\hat{N}_1\rangle$ as in Eq.~\eqref{e:Neta1}, but with $m$ replaced by $|\mu|^2$ (after setting $n=0$). 

\section{Phase sensitivity when seeding with a number state}\label{a:phasesensitivity_number}
\noindent We calculate the phase sensitivity $\Delta\phi^2$ by obtaining the numerator and denominator in Eq.~\eqref{e:Deltaphisq} separately. The numerator of Eq.~\eqref{e:Deltaphisq} is the variance of $\hat{N}_1$, given by $\langle\hat{N}_1^2\rangle-\langle\hat{N}_1\rangle^2$, which we obtain as follows. First, we calculate $\langle\hat{N}_1^2\rangle$ by computing $\hat{b}_1^\dagger\hat{b}_1\ket{m}_{a_2}$ by means of the overall transformation matrix in Eq.~\eqref{e:overallmatrix},
\begin{align*}
	\hat{b}_1^\dagger\hat{b}_1\ket{m}_{a_2}=&\
	\alpha_1\sqrt{m+1}\left(A_1^*\ket{1}_{a_1}\ket{m+1}_{a_2}+\alpha_1^*\sqrt{m+1}\ket{m}_{a_2}+B_1^*\ket{m+1}_{a_2}\ket{1}_{\ell_1}+\sqrt{1-\eta_1}\ket{m+1}_{a_2}\ket{1}_{d_1}\right)\\
	&+\beta_1\left(A_1^*\ket{1}_{a_1}\ket{m}_{a_2}\ket{1}_{\ell_2}+\alpha_1^*\sqrt{m}\ket{m-1}_{a_2}\ket{1}_{\ell_2}+B_1^*\ket{m}_{a_2}\ket{1}_{\ell_1}\ket{1}_{\ell_2}+\beta_1^*\ket{m}_{a_2}+\sqrt{1-\eta_1}\ket{m}_{a_2}\ket{1}_{\ell_2}\ket{1}_{d_1}\right)\\
	=&\ \left[|\alpha_1|^2(m+1)+|\beta_1|^2\right]\ket{m}_{a_2}+\alpha_1A_1^*\sqrt{m+1}\ket{1}_{a_1}\ket{m+1}_{a_2}+\alpha_1B_1^*\sqrt{m+1}\ket{m+1}_{a_2}\ket{1}_{\ell_1}\\
	&+\alpha_1\sqrt{m+1}\sqrt{1-\eta_1}\ket{m+1}_{a_2}\ket{1}_{d_1}+\beta_1A_1^*\ket{1}_{a_1}\ket{m}_{a_2}\ket{1}_{\ell_2}+\beta_1\alpha_1^*\sqrt{m}\ket{m-1}_{a_2}\ket{1}_{\ell_2}+\beta_1B_1^*\ket{m}_{a_2}\ket{1}_{\ell_1}\ket{1}_{\ell_2}\\
	&+\beta_1\sqrt{1-\eta_1}\ket{m}_{a_2}\ket{1}_{\ell_2}\ket{1}_{d_1}.
\end{align*}
Then, we take the Hermitian conjugate of $\hat{b}_1^\dagger\hat{b}_1\ket{m}_{a_2}$ and calculate the inner product 
\begin{align*}
	\langle\hat{N}_1^2\rangle=\bra{m}_{a_2}\hat{b}_1^\dagger\hat{b}_1\hat{b}_1^\dagger\hat{b}_1\ket{m}_{a_2}.
\end{align*}
The result is
\begin{align}
	\langle\hat{N}_1^2\rangle=&\ \left[|\alpha_1|^2(m+1)+|\beta_1|^2\right]^2+|A_1|^2|\alpha_1|^2(m+1)+|\alpha_1|^2|B_1|^2(m+1)\notag\\
	&+|\alpha_1|^2(m+1)(1-\eta_1)+|A_1|^2|\beta_1|^2+m|\alpha_1|^2|\beta_1|^2+|B_1|^2|\beta_1|^2\label{e:Neta1sq}\\
	&+|\beta_1|^2(1-\eta_1).\notag
\end{align}
Therefore, 
\begin{align}
	\Delta\hat{N}_1^2=&\ \langle\hat{N}_1^2\rangle-\langle\hat{N}_1\rangle^2\notag\\
	=&\ |A_1|^2|\alpha_1|^2(m+1)+|\alpha_1|^2|B_1|^2(m+1)\notag\\
	&+|\alpha_1|^2(m+1)(1-\eta_1)+|A_1|^2|\beta_1|^2+m|\alpha_1|^2|\beta_1|^2+|B_1|^2|\beta_1|^2\notag\\
	&+|\beta_1|^2(1-\eta_1)\notag\\
	=&\ |\alpha_1|^2(m+1)\left(|A_1|^2+|B_1|^2+1-\eta_1\right)+|\beta_1|^2\left(|A_1|^2+m|\alpha_1|^2+|B_1|^2+1-\eta_1\right),\label{e:VarNeta1}
\end{align}
where we used
\[
	\langle\hat{N}_1\rangle^2=\left[|\alpha_1|^2(m+1)+|\beta_1|^2\right]^2,
\]
from Eq.~\eqref{e:nmbbnm} after setting $n=0$. Note that although not written explicitly, $\Delta\hat{N}_1^2$ depends on $\phi$ via $|A_1|^2$ and $|\alpha_1|^2$ in Eqs.~\eqref{e:A1sq} and~\eqref{e:alpha1sq}, respectively. Finally, we compute the denominator of Eq.~\eqref{e:Deltaphisq}, which contains the derivative $\partial\langle\hat{N}_1\rangle/\partial\phi$, with $\langle\hat{N}_1\rangle$ given by Eq.~\eqref{e:Neta1}. This derivative is equal to
\begin{equation}
	\frac{\partial\langle\hat{N}_1\rangle}{\partial\phi}=-2\eta_1(m+1)\sqrt{T_1T_2U_AU_BV_AV_B}\sin\phi.\label{e:dNdphi}
\end{equation}
By combining the results in Eqs.~\eqref{e:VarNeta1} and~\eqref{e:dNdphi}, we finally obtain $\Delta\phi^2$ in Eq.~\eqref{e:Deltaphisq_number}.

\section{Phase sensitivity when seeding with a coherent state}\label{a:phasesensitivity_coherent}
\noindent Let us calculate $\Delta\hat{N}_1^2$ when seeding mode $a_2$ with a coherent state $\ket{\mu}_{a_2}$. Since we already know $\langle\hat{N}_1\rangle$ from Eq.~\eqref{e:Neta1_coherent}, we just need to calculate $\langle\hat{N}_1^2\rangle$. Following the same steps as in Appendix~\ref{a:phasesensitivity_number}, we first obtain $\hat{b}^\dagger_1\hat{b}_1\ket{\mu}_{a_2}$,
\begin{align*}
	\hat{b}_1^\dagger\hat{b}_1\ket{\mu}_{a_2}=&\ \alpha_1e^{-\frac{|\mu|^2}{2}}\sum_{q=0}^\infty\frac{\mu^q}{\sqrt{q!}}\sqrt{q+1}\left(A_1^*|1\rangle_{a_1}|q+1\rangle_{a_2}+\alpha_1^*\sqrt{q+1}\ket{q}_{a_2}+B_1^*\ket{q+1}_{a_2}\ket{1}_{\ell_1}+\sqrt{1-\eta_1}\ket{q+1}_{a_2}\ket{1}_{d_1}\right)\\
	&+\beta_1\bigg(A_1^*|1\rangle_{a_1}|\mu\rangle_{a_2}|1\rangle_{\ell_2}+\alpha_1^*e^{-\frac{|\mu|^2}{2}}\sum_{q=0}^\infty\frac{\mu^q}{\sqrt{q!}}\sqrt{q}\ket{q-1}_{a_2}\ket{1}_{\ell_2}+B_1^*\ket{\mu}_{a_2}\ket{1}_{\ell_1}\ket{1}_{\ell_2}+\beta_1^*\ket{\mu}_{a_2}\\
	&\phantom{+\beta_1\bigg(}+\sqrt{1-\eta_1}\ket{\mu}_{a_2}\ket{1}_{\ell_2}\ket{1}_{d_1}\bigg)\\
	=&\ e^{-\frac{|\mu|^2}{2}}\sum_{q=0}^\infty\frac{\mu^q}{\sqrt{q!}}\Big\{\left[|\alpha_1|^2(q+1)+|\beta_1|^2\right]\ket{q}_{a_2}+\alpha_1A_1^*\sqrt{q+1}\ket{1}_{a_1}\ket{q+1}_{a_2}+\alpha_1B_1^*\sqrt{q+1}\ket{q+1}_{a_2}\ket{1}_{\ell_1}\\
	&\phantom{e^{-\frac{|\mu|^2}{2}}\sum_{q=0}^\infty\frac{\mu^q}{\sqrt{q!}}\Big\{}+\alpha_1\sqrt{q+1}\sqrt{1-\eta_1}\ket{q+1}_{a_2}\ket{1}_{d_1}+\beta_1A_1^*\ket{1}_{a_1}\ket{q}_{a_2}\ket{1}_{\ell_2}+\beta_1\alpha_1^*\sqrt{q}\ket{q-1}_{a_2}\ket{1}_{\ell_2}\\
	&\phantom{e^{-\frac{|\mu|^2}{2}}\sum_{q=0}^\infty\frac{\mu^q}{\sqrt{q!}}\Big\{}+\beta_1B_1^*\ket{q}_{a_2}\ket{1}_{\ell_1}\ket{1}_{\ell_2}+\beta_1\sqrt{1-\eta_1}\ket{q}_{a_2}\ket{1}_{\ell_2}\ket{1}_{d_1}\Big\}.
\end{align*}
We then calculate the Hermitian conjugate of $\hat{b}_1^\dagger\hat{b}_1\ket{\mu}_{a_2}$ and take the inner product 
\begin{equation*}
	\langle\hat{N}_1^2\rangle=\bra{\mu}_{a_2}\hat{b}_1^\dagger\hat{b}_1\hat{b}_1^\dagger\hat{b}_1\ket{\mu}_{a_2},
\end{equation*}
to find
\begin{align*}
	\langle\hat{N}_1^2\rangle=&\ e^{-|\mu|^2}\sum_{q=0}^\infty\frac{|\mu|^{2q}}{q!}\Big\{\left[|\alpha_1|^2(q+1)+|\beta_1|^2\right]^2+|A_1|^2|\alpha_1|^2(q+1)+|\alpha_1|^2|B_1|^2(q+1)\\
	&\phantom{e^{-|\mu|^2}\sum_{q=0}^\infty\frac{|\mu|^{2q}}{q!}\Big\{}+|\alpha_1|^2(q+1)(1-\eta_1)+|A_1|^2|\beta_1|^2+q|\alpha_1|^2|\beta_1|^2\\
	&\phantom{e^{-|\mu|^2}\sum_{q=0}^\infty\frac{|\mu|^{2q}}{q!}\Big\{}+|B_1|^2|\beta_1|^2+|\beta_1|^2(1-\eta_1)\Big\}\\
	=&\ e^{-|\mu|^2}\sum_{q=0}^\infty\frac{|\mu|^{2q}}{q!}\left[|\alpha_1|^2(q+1)+|\beta_1|^2\right]^2+|A_1|^2|\alpha_1|^2(|\mu|^2+1)+|\alpha_1|^2|B_1|^2(|\mu|^2+1)\\
	&+|\alpha_1|^2(|\mu|^2+1)(1-\eta_1)+|A_1|^2|\beta_1|^2+|\mu|^2|\alpha_1|^2|\beta_1|^2\\
	&+|B_1|^2|\beta_1|^2+|\beta_1|^2(1-\eta_1).
\end{align*}
So far, the expression for $\langle\hat{N}_1^2\rangle$ looks like in Eq.~\eqref{e:Neta1sq} after replacing $m$ by $|\mu|^2$, respectively, except for the term in square brackets. Let us take a closer look at this term,
\begin{align*}
	e^{-|\mu|^2}\sum_{q=0}^\infty\frac{|\mu|^{2q}}{q!}\left[|\alpha_1|^2(q+1)+|\beta_1|^2\right]^2=&\
	e^{-|\mu|^2}\sum_{q=0}^\infty\frac{|\mu|^{2q}}{q!}\left[|\alpha_1|^4(q^2+2q+1)+2|\alpha_1|^2|\beta_1|^2(q+1)+|\beta_1|^4\right]\\
	=&\ |\alpha_1|^4(|\mu|^4+3|\mu|^2+1)+2|\alpha_1|^2|\beta_1|^2(|\mu|^2+1)+|\beta_1|^4\\
	=&\ \left[|\alpha_1|^2(|\mu|^2+1)+|\beta_1|^2\right]^2+|\mu|^2|\alpha_1|^4,
\end{align*}
where we used the identity
\begin{gather*}
    e^{-|\mu|^2}\sum_{q=0}^\infty\frac{|\mu|^{2q}}{q!}q^2=|\mu|^4+|\mu|^2.
\end{gather*}
We observe that the term in square brackets in $\langle\hat{N}_1^2\rangle$ is not the same as the one in Eq.~\eqref{e:Neta1sq} after substituting $m$ by $|\mu|^2$. Instead, there is an additional term $|\mu|^2|\alpha_1|^4$, resulting from the fact that the mode $a_2$ is initially in the coherent state $\ket{\mu}_{a_2}$. Thus, $\langle\hat{N}_1^2\rangle$ becomes
\begin{align*}
	\langle\hat{N}_1^2\rangle=&\ \left[|\alpha_1|^2(|\mu|^2+1)+|\beta_1|^2\right]^2+|\mu|^2|\alpha_1|^4+|A_1|^2|\alpha_1|^2(|\mu|^2+1)+|\alpha_1|^2|B_1|^2(|\mu|^2+1)\\
	&+|\alpha_1|^2(|\mu|^2+1)(1-\eta_1)+|A_1|^2|\beta_1|^2+|\mu|^2|\alpha_1|^2|\beta_1|^2\\
	&+|B_1|^2|\beta_1|^2+|\beta_1|^2(1-\eta_1),
\end{align*}
and $\Delta\hat{N}_1^2$ reduces to
\begin{align}
	\Delta\hat{N}_1^2=&\ \langle\hat{N}_1^2\rangle-\langle\hat{N}_1\rangle^2\notag\\
	=&\ |\mu|^2|\alpha_1|^4+|A_1|^2|\alpha_1|^2(|\mu|^2+1)+|\alpha_1|^2|B_1|^2(|\mu|^2+1)\notag\\
	&+|\alpha_1|^2(|\mu|^2+1)(1-\eta_1)+|A_1|^2|\beta_1|^2+|\mu|^2|\alpha_1|^2|\beta_1|^2\notag\\
	&+|B_1|^2|\beta_1|^2+|\beta_1|^2(1-\eta_1)\notag\\
	=&\ |\mu|^2|\alpha_1|^4+|\alpha_1|^2(|\mu|^2+1)\left(|A_1|^2+|B_1|^2+1-\eta_1\right)+|\beta_1|^2\left(|A_1|^2+|\mu|^2|\alpha_1|^2+|B_1|^2+1-\eta_1\right),\label{e:VarNeta1_coherent}
\end{align}
where we recalled Eq.~\eqref{e:Neta1_coherent} to calculate $\langle\hat{N}_1\rangle^2$. Similar as before, although not written explicitly, $\Delta\hat{N}_1^2$ depends on $\phi$ via $|A_1|^2$ and $|\alpha_1|^2$.

To finally obtain the phase sensitivity, we derive the result in Eq.~\eqref{e:VarNeta1_coherent} by the derivative of $\langle\hat{N}_1\rangle$ in Eq.~\eqref{e:Neta1coh} with respect to $\phi$. The resulting derivative is the same as in Eq.~\eqref{e:dNdphi} after substituting $m$ by $|\mu|^2$. Alternatively, the results for $\Delta\phi^2$ when seeding with coherent states can be derived via the Gaussian formalism~\cite{Olivares12,Vallone19}, which does not involve explicit series sums.
\end{widetext}

\bibliography{EnhancedNonlinearInterferometryViaSeeding}

\end{document}